\documentclass[manuscript]{aastex}

\usepackage{xspace}
\usepackage{amssymb}
\usepackage{times}
\usepackage{verbatim}
\usepackage{color}
\usepackage{natbib}

\shorttitle{Solar Flare Plasma Composition}
\shortauthors{Sylwester et al.}

\begin{document}


\title{Solar Flare Composition and Thermodynamics from RESIK X-ray Spectra}


\author{B. Sylwester\altaffilmark{1} and  J. Sylwester\altaffilmark{1}}
\affil{$^1$ Space Research Center, Polish Academy of Sciences, Kopernika 11, 51-622 Wroc{\l}aw, Poland}
\email{bs,js@cbk.pan.wroc.pl}
\author{K. J. H. Phillips\altaffilmark{2}}
\affil{$^2$ Earth Sciences Department, Natural History Museum, London SW7 5BD, United Kingdom}
\email{kennethjhphillips@yahoo.com}
\author{A. K\c{e}pa\altaffilmark{1} and T. Mrozek \altaffilmark{1,3}}
\affil{$^3$ Astronomical Institute, University of Wroc{\l}aw, ul. Kopernika 11, 51-622 Wroc{\l}aw, Poland}
\email{ak,tmrozek @cbk.pan.wroc.pl}

%\date{\today}

\altaffiltext{2}{Scientific Associate}


\begin{abstract}

Previous estimates of the solar flare abundances of Si, S, Cl, Ar, and K from the RESIK X-ray crystal spectrometer on board the {\em CORONAS-F} spacecraft were made on the assumption of isothermal X-ray emission. We investigate the effect on these estimates by relaxing this assumption and instead determining the differential emission measure (DEM) or thermal structure of the emitting plasma by re-analyzing RESIK data for a {\em GOES} class M1.0 flare on 2002 November~14 (SOL2002-11-14T22:26) for which there was good data coverage. The analysis method uses a maximum-likelihood (Withbroe--Sylwester) routine for evaluating the DEM. In a first step, called here AbuOpt, an optimized set of abundances of Si, S, Ar, and K is found that is consistent with the observed spectra. With these abundances, the differential emission measure evolution during the flare is found. The abundance optimization leads to revised abundances of silicon and sulfur in the flare plasma: $A({\rm S}) = 6.94 \pm 0.06$ and $A({\rm Si}) = 7.56 \pm 0.08$ (on a logarithmic scale with $A({\rm H}) = 12$). Previously determined abundances of Ar, K, and Cl from an isothermal assumption are still the preferred values. During the flare's maximum phase, the X-ray-emitting plasma has a basically two-temperature structure, with the cooler plasma with approximately constant temperature (3--6~MK) and a hotter plasma with temperature $16-21$~MK. Using imaging data from the {\em RHESSI} hard X-ray spacecraft, the emission volume of the hot plasma is deduced from which lower limits of the electron density $N_e$ and the thermal content of the plasma are given.
\end{abstract}


% Section 1
\section{Introduction}

Observations of solar soft X-ray spectra are essential for the diagnostics of hot plasmas associated with flares and active regions. The fluxes of emission lines and continua depend sensitively on electron temperature or rather, since the plasma is not in general isothermal, the distribution of emission measure with electron temperature. For rapidly varying conditions, X-ray spectra can also be used to find the plasma's ionization state or the presence of nonthermal electrons. If combined with images taken at similar energy ranges, lower limits of electron densities and the energy content of the emitting plasma can also be found.

The RESIK (REntgenovsky Spektrometr s Izognutymi Kristalami: \citet{jsyl05}) instrument, a crystal spectrometer aboard the Russian {\it CORONAS-F} spacecraft, is one of several such spectrometers over the past 30 years or so using a bent crystal geometry with position-sensitive detectors, so that complete spectra over particular ranges can be captured in short time intervals and with much higher sensitivity than is possible with flat scanning spectrometers. RESIK obtained spectra from non-flaring active regions and during numerous flares between 2001 and 2003, the only crystal spectrometer to do so. This period occurred during the latter part of Cycle~23, when the activity levels were higher than at any time since, including the peak of the present cycle (number 24). Consequently, a number of large active regions and flares were observed, and the results already discussed; a catalog of observations is available at http://www.cbk.pan.wroc.pl/experiments/resik/resik\_catalogue.htm. Four spectral bands covered the nominal range 3.4~\AA--6.1~\AA\ with two silicon crystals (Si 111, $2d = 6.27$~\AA) for channels 1 and 2 (spectral ranges for an on-line source 3.40~\AA--3.80~\AA, 3.83~\AA--4.27~\AA) and two quartz crystals (quartz $10\bar 10$, $2d = 8.51$~\AA) for channels 3 and 4 (spectral ranges 4.35~\AA--4.86~\AA, 5.00~\AA--6.05~\AA). Instrumental fluorescence background emission, often a problem with previous solar X-ray spectrometers, was minimized through the adjustment of pulse-height analyzer settings over the mission lifetime; ultimately, for the period 2002 December~24 to 2003 March~23, the fluorescence background was entirely eliminated for channels 1 and 2 and its amount reduced and accurately estimated for channels 3 and 4. To maximize the instrument's sensitivity, no collimator was used; although this introduced the possibility of overlapping spectra from two or more X-ray sources on the Sun, in practice this very rarely occurred.

RESIK was intensity-calibrated to a higher accuracy than was possible for previous solar crystal spectrometers (the procedure is described by \citet{jsyl05}), so enabling element abundances to be derived for elements whose spectral lines feature in RESIK spectra. Previously, such analyses (see e.g. \citet{jsyl10a}) have used the assumption of an isothermal plasma for the X-ray emission and with temperature and emission measure given by the flux ratio of the two emission bands of {\em GOES}. The justification for this was that plots of the measured line fluxes during flares divided by the {\em GOES} emission measure ($EM_{\rm GOES}$) against {\em GOES} temperature ($T_{\rm GOES}$) showed points distributed either along the theoretical $G(T_e)$ function or the function displaced by a constant amount. The $G(T_e)$ function is the line emission per unit emission measure as a function of electron temperature $T_e$ calculated, e.g., from the {\sc chianti} atomic database and software package \citep{der97,lan12} for an assumed element abundance. The amount of the displacement gives the factor by which the assumed abundance must be adjusted to give agreement with the measured RESIK line fluxes. A particularly tight distribution of points around the calculated $G(T_e)$ curve was obtained for the case of the \ion{Ar}{17} lines in RESIK's channel~2, and a rather broader scatter of points for K and Cl since the line emission for these low-abundance elements was weak. Thus an argon abundance estimate with very small statistical uncertainty ($A({\rm Ar}) = 6.45 \pm 0.06$ on a logarithmic scale with $A({\rm H}) = 12$) resulted, in close agreement with other argon abundance estimates from solar proxies \citep{jsyl10b}. The RESIK Si and S abundance estimates are based on strong lines of H-like and He-like Si and S seen in RESIK's channels 3 and 4, but the distribution of points given by line flux divided by $EM_{\rm GOES}$ against $T_{\rm GOES}$ was less impressive than that for the \ion{Ar}{17} lines. It was speculated \citep{jsyl12,bsyl13} that the subtraction of crystal fluorescence was not as accurately done as was thought or that the results were affected by some lines occurring very near the end of the range of either channel 3 or channel 4. Here we investigate whether the assumption of an isothermal emitting plasma might instead be more significant in leading to biased results, the thinking being that the temperature derived from  the two {\em GOES} channels is more representative of the hotter \ion{Ar}{17} lines than that of the H-like or He-like Si and S ions, an idea described further in Section~3.

In this work, we discuss RESIK spectra for the particular case of the M1.0 flare on 2002 November~14 with {\em GOES} soft X-ray maximum at 22:26~UT (SOL2002-11-14T22:26 using the IAU standard flare-naming convention). We first used an iterative procedure (AbuOpt) to derive optimized element abundance estimates of Si, S, Ar, and K, since these elements are represented by lines in RESIK spectra and so their abundances will influence the nature of the RESIK spectra. The optimization is done with a maximum likelihood routine (the Withbroe--Sylwester routine) which determines the differential emission measure (DEM). The optimized abundances of Si and S in particular differ from our previous estimates based on an isothermal assumption. With the Withbroe--Sylwester routine, and with the optimized abundances of Si, S, Ar, and K, the evolution of the differential emission measure (DEM) over the flare duration was then found. With X-ray images of this flare from {\it RHESSI}, estimates of the emitting volume $V$ are made, and from these, lower limits to the electron densities and thermal energy content of the flaring plasma are determined. The physical significance of the new Si and S abundances are discussed in respect of the well known first ionization potential (FIP) effect, in which the abundances of elements with low ($\lesssim 10$~eV) FIP in coronal plasmas are apparently enhanced over photospheric abundances.

% Section 2
\section{Observations}

\begin{figure}
\epsscale{.80}
\plotone{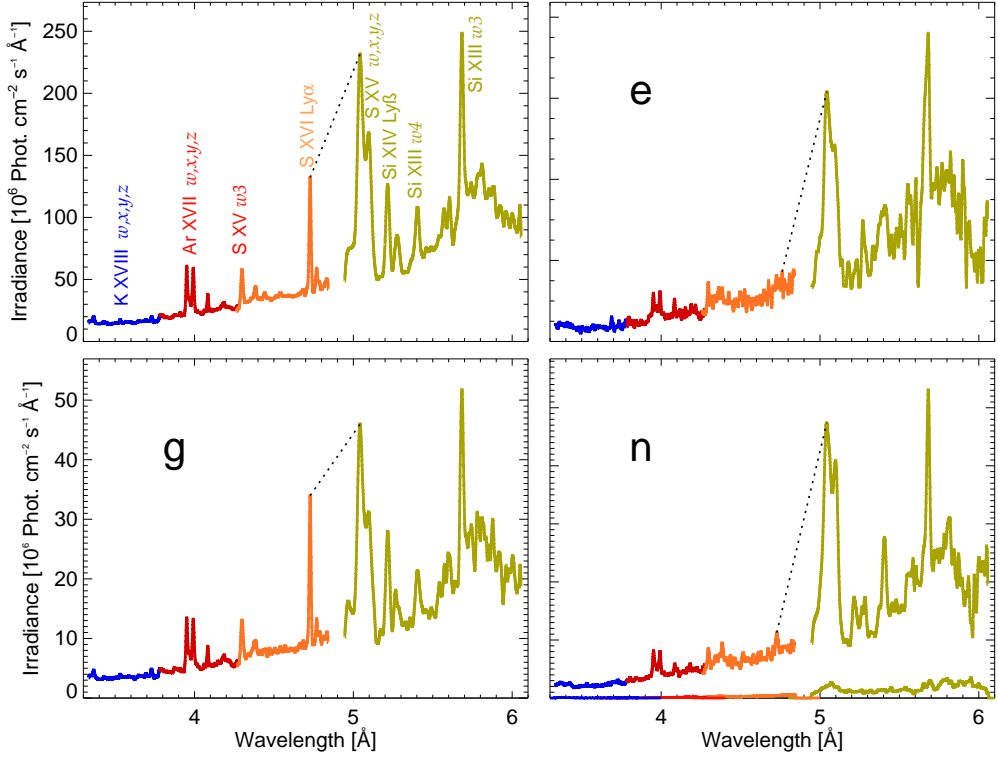}
\caption{(Top left:) Averaged RESIK spectrum for the SOL2002-11-14T22:26 flare over the period 22:22:45--22:53:31~UT (total exposure time 26.5 minutes). Different colors are used for the four RESIK channels. Principal line identifications of the chief features are given; the notation used is for the He-like ions (\ion{K}{18}, \ion{S}{15}, \ion{Si}{13}) are $w = 1s^2\,^1S_0 - 1s2p\,^1P_1$, $w3 = 1s^2\,^1S_0 - 1s3p\,^1P_1$, $w4 = 1s^2\,^1S_0 - 1s4p\,^1P_1$. (Top right and lower panels:) Three representative spectra taken during the rise (22:22:46--22:24:48~UT, upper right $e$), maximum (22:25:46--22:26:46~UT, lower left $g$), and decay (22:34:06--22:36:36, lower right $n$) phases of the flare, with integration times 2, 1, and 2.5~minutes (the letters refer to the intervals shown in Figure~\ref{norm_RHESSI_GOES_lcs}). The pre-flare spectrum is shown near the zero level of panel $n$. The inclination of the dotted track linking the \ion{S}{15} (5.04~\AA) and the hotter \ion{S}{16} (4.73~\AA) line features is an indication of the emitting plasma's changing temperature.}
\label{RESIK_spectra}
\end{figure}

The SOL2002-11-14T22:26 flare occurred in active region NOAA~10195 located at S14E65, a region that had recently rotated on to the south-east limb and became flare-productive on 2012 November~14. During the November~14 flare under discussion, the X-ray emission, indicated by both RESIK and {\em GOES}, showed a sudden enhancement starting at 22:15~UT with maximum at about 22:26~UT, followed by a slow decay extending until spacecraft night, at 22:56~UT. Some 124 individual RESIK spectra (total integrated time 26.5~minutes) are included in the spectrum shown in Figure~\ref{RESIK_spectra} (top left panel) which is averaged over the entire flare duration. The top right and lower panels ($e$, $g$, $n$) show RESIK spectra averaged over time intervals (given in the caption) during the rise, maximum, and decay phases of the flare. The changing relative line fluxes of the \ion{S}{15} lines near 5.04~\AA\ (channel~4) and \ion{S}{16} Ly-$\alpha$ line at 4.73~\AA, connected by dotted lines, are evident, indicating as expected that the maximum temperature is attained near the flare peak phase. Three RESIK spectra were taken during the previous spacecraft orbit (from 21:15:16~UT to 21:18:52~UT) when no soft X-ray flare was in progress and the total X-ray activity was much lower (\emph{GOES} C1 level); the average of these spectra is shown near the zero level of panel $n$, clearly showing that the spectra shown in Figure~\ref{RESIK_spectra} are almost entirely due to the flare. The chief spectral line features are due to transitions in H-like and He-like ions of Si, S, Ar, and K and are formed by thermal plasmas with temperature range 2~MK to 20~MK. Among those of note are the He-like K (\ion{K}{18} $1s^2-1s2s, 1s2p$) lines (see \citet{jsyl10a}) and H-like and He-like Ar lines in channel~1; the He-like Ar (\ion{Ar}{17} $1s^2-1s2s, 1s2p$) lines  \citep{jsyl10b} and the He-like S (\ion{S}{15}) $1s^2-1s4p$ line in channel~2; the \ion{S}{15} $1s^2-1s3p$ and H-like S (\ion{S}{16}) Ly-$\alpha$ lines \citep{jsyl12} and He-like Cl (\ion{Cl}{17}) lines \citep{bsyl11} in channel~3; and \ion{S}{15} $1s^2-1s2s, 1s2p$ lines and H-like Si (\ion{Si}{14}) and He-like Si (\ion{Si}{13}) lines and some Si~{\sc xii} dielectronic satellites \citep{jsyl12,bsyl13} in channel~4. Further details about these lines including wavelengths are given in the papers cited and in Table~II of \citet{jsyl05}.

%%%%%%%%%%%%%%%%%%%%%%%%%%%%%%%  Figure 1   %%%%%%%%%%%%%%%%%%%%%%%%%%%%%%%%%%%%%%%%%%%%%%%%%%%%%%%
%%%%%%%%%%%%%%%%%%%%%%%%%%%%%%%%%%%  Figure 1  %%%%%%%%%%%%%%%%%%%%%%%%%%%%%%%%%%%%%%%%%%%%%%%%

%%%%%%%%%%%%%%%%%%%%%%%%%%%%%%%  Figure 2  %%%%%%%%%%%%%%%%%%%%%%%%%%%%%%%%%%%%%%%%%%%%%%%%%%%%%%%
\begin{figure}
\epsscale{.80}
\plotone{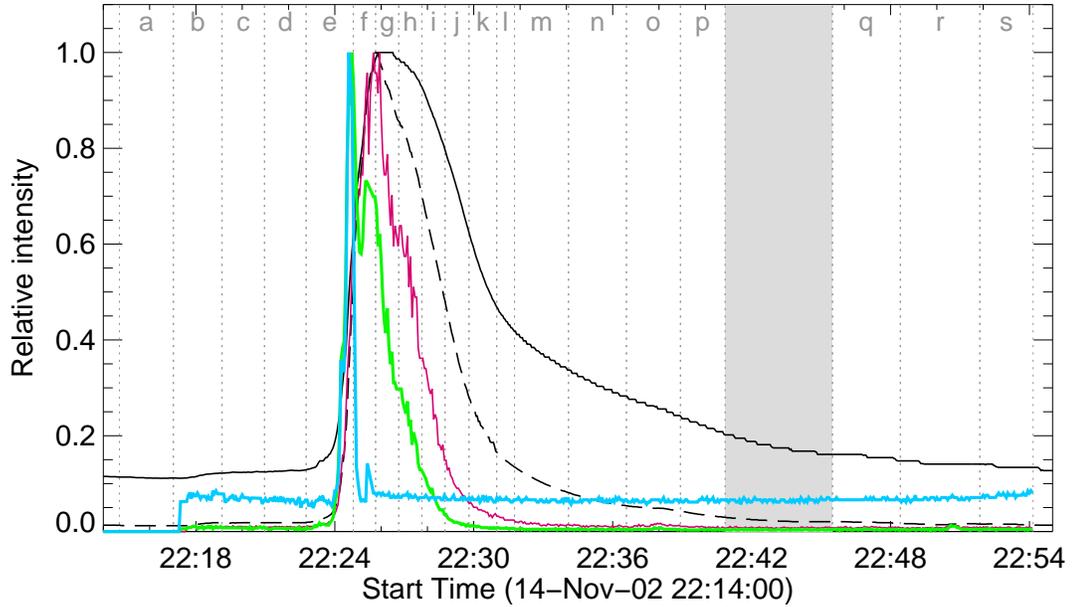}
\vspace{6mm}
\caption{Normalized {\em RHESSI} and {\em GOES} light curves for the SOL2002-11-14T22:26 flare. The \emph{GOES} fluxes in the 1~--~8~\AA\ and 0.5~--~4~\AA\ ranges are shown as black continuous and dashed lines respectively. The \emph{RHESSI} light curves are shown by colored lines: pink (6~--~12~keV), green (12~--~25~keV), and blue (25~--~50~keV). The key letters above each time strip denote intervals over which RESIK spectra were integrated for DEM analysis. The grey strip indicates a passage through a polar van Allen radiation belt when the RESIK high-voltages were turned off and no observations were made.}
\label{norm_RHESSI_GOES_lcs}
\end{figure}
%%%%%%%%%%%%%%%%%%%%%%%%%%%%%%%%%%%  Figure 2  %%%%%%%%%%%%%%%%%%%%%%%%%%%%%%%%%%%%%%%%%%%%%%%%

%%%%%%%%%%%%%%%%%%%%%%%%%%%%%%%  Figure 3   %%%%%%%%%%%%%%%%%%%%%%%%%%%%%%%%%%%%%%%%%%%%%%%%%%%%%%%
\begin{figure}
\epsscale{.60}
\plotone{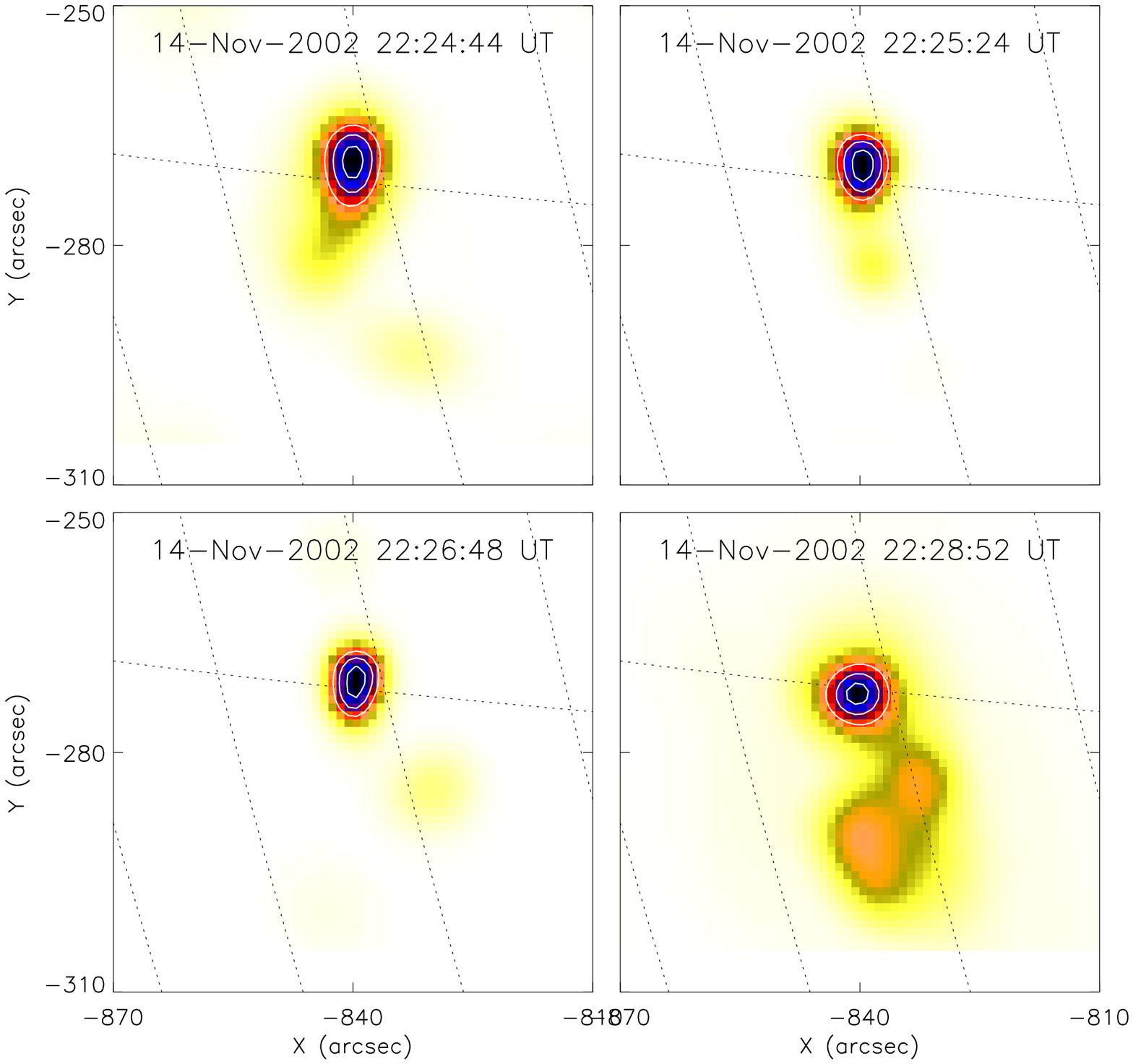}
\caption{Four \emph{RHESSI} images of SOL2002-11-14T22:26 flare obtained with the PIXON algorithm in the energy range 6~--~7~keV. Times (UT) are indicated at the top of each panel. The top two images were taken during hard X-ray impulsive emission, the lower left image at the soft X-ray maximum, and the lower right image during the decay phase. The contours are drawn at levels 0.5, 0.7 and 0.9 of the maximum intensity in each image. Solar north is at the top of each image, solar east to the left; the $x$ (east--west) and $y$ (north--south) co-ordinates are in arcseconds. The dotted grid lines show solar longitude and latitude at $3^\circ$ intervals. The integration time for the last image was 12~s, the first three images 4~s (i.e. one rotation of the {\em RHESSI} spacecraft). The low-intensity feature to the south of the main feature in the last image is reproducible and so is likely to be real, but its weakness does not significantly add to the total emission. The extent of the main flare emission at the 50\% iso-contour level is approximately 5400~km. }
\label{RHESSI_images}
\end{figure}
%%%%%%%%%%%%%%%%%%%%%%%%%%%%%%%%%%%  Figure 3  %%%%%%%%%%%%%%%%%%%%%%%%%%%%%%%%%%%%%%%%%%%%%%%%

Figure~\ref{norm_RHESSI_GOES_lcs} shows the normalized X-ray light curves for the flare in {\em RHESSI} and {\em GOES} with color codes for different energy bands (the two {\em GOES} bands and three selected from available {\em RHESSI} bands). The harder X-rays seen with {\em RHESSI} peak early in the flare development, with the 12~--~25~keV and 25~--~50~keV emission showing sharp impulsive bursts at 22:24:42~UT and 22:25:27~UT, during the soft X-ray emission rise seen with {\em GOES}. This figure also shows 19 time intervals (marked by key letters $a$ to $s$) over which RESIK spectra were integrated and used for further analysis, in particular the inversion of line fluxes to obtain differential emission measure for which spectra of high statistical quality are necessary. (Letters $e$, $g$, and $n$ in the panels of Figure~\ref{RESIK_spectra} refer to those shown in Figure~\ref{norm_RHESSI_GOES_lcs}.) Figure~\ref{RHESSI_images} shows {\em RHESSI} images of the flare, with the upper two images were taken during the first and second hard impulsive X-ray peaks, the lower two images during the soft X-ray maximum (22:26:48~UT) and the decay phase (22:28:52~UT). The image reconstructions were performed using the PIXON method and {\em RHESSI} grids 3, 4, 5, 6, 8, and 9; this was done for the 6~--~7~keV energy range, which is effectively the lowest accessible to {\em RHESSI} and so is larger than the energy range of the RESIK spectra (2.0--3.6~keV). In the analysis below, we are interested in the {\em RHESSI} image dimensions so the relative merits of image construction routines are of some importance. These are discussed by \cite{den09}. The PIXON method is the only way to image extended sources in the presence of compact sources, as in the case here, and there is much less background compared with the CLEAN routine which is also available. There is the possibility of ``over-resolution" (i.e. the source size is under-represented) with PIXON, but only if the total signal in the image exceeds 6000--7000 photon counts: in the case of the images shown in Figure~\ref{RHESSI_images}, the total signal is slightly less than 2000 counts. The contours on the images in Figure~\ref{RHESSI_images} are drawn at levels 0.5, 0.7 and 0.9 of the maximum emission in each image. Approximately 60 such images over the flare duration show an additional weak feature to the south of the main structure; this does not add to estimates of the image extent which, at the 0.5 contour level, is almost constant at 5400~km.

% Section 3
\section{Spectral analysis method}

A commonly used method of deducing electron temperature $T_e$ and volume emission measure $EM = \int N_e^2 dV$ for X-ray flare emission assumed to be isothermal uses the flux ratio of the {\em GOES} 1--8~\AA\ and 0.5--4.0~\AA\ bands based on the work of \citet{white05}. However, an isothermal assumption is not generally valid for the interpretation of X-ray spectra with many lines formed over a broad temperature range, different temperatures being obtained according to the flux ratio of the lines chosen. More generally, the emission measure within intervals of temperature is described by the differential emission measure, DEM \citep{with75,lev80,mct99,landi08}. Thus, for an optically thin, multi-thermal plasma such as the X-ray flare discussed here, the observed flux $F_i$ of each line or spectral interval \emph{i} can be expressed as \citep{jsyl80}:

% Eq. 1
\begin{equation}
F_i = A_i \int_{T=0}^{\infty} f_i(T)\varphi(T) {\rm d}T
\label{lineflux}
\end{equation}

\noindent where $A_i$ represents the assumed abundance of an element contributing to the flux of a particular line or spectral interval. We assume  $A_i$ to be constant over emitting volume $V$, while $f_i(T)$, the emission function (more commonly known as the $G(T)$ function for individual lines) is calculable from atomic excitation theory for chosen spectral intervals $i$. The differential emission measure function $\varphi(T)$, defined by

% Eq. 2
\begin{equation}
DEM \equiv \varphi(T) \equiv {N_e}^2  \frac{{\rm d}V}{{\rm d}T}\,\,,
\end{equation}

\noindent may be determined by solving under certain conditions from the analysis of a full system of equations like Equation~(\ref{lineflux}) for $i$ spectral intervals. Although the problem is recognized as being ill-conditioned \citep{cra76}, methods of solution exist that give solutions for $\varphi(T)$ and their uncertainties that satisfy the input data to within observational uncertainties and can be ascribed physical meaning. Here, we followed the iterative maximum-likelihood, Bayesian routine called the Withbroe--Sylwester method and described by \citet{jsyl80}. It has been tested previously on synthetic spectra and assumed DEM functions to see whether the DEMs are recovered after the inversion (e.g. \citet{jsyl98,jsyl99,kepa06}).

% Table 1: List of RESIK spectral intervals
\begin{table}[h]
\caption{Wavelength bands used for studies of the elemental abundances and DEM}  % title of Table
\vspace{5mm}
\label{table:1}      % is used to refer this table in the text
\centering                          % used for centering table
\begin{tabular}{c c c}        % centered columns (4 columns)
\hline\hline                 % inserts double horizontal lines
No & Range [\AA] & Main contributor (cont. = continuum) \\    % table heading
\hline                        % inserts single horizontal line
\vspace{-3mm}
     &               &                                                       \\
    1 & 3.480~-~3.630 & cont. + K~{\sc xviii }$2p$+sat.                      \\
    2 & 3.630~-~3.800 & cont. + Ar~{\sc xviii }$2p$, S~{\sc xvi }$4p, 5p$    \\
    3 & 3.900~-~4.060 & cont. + Ar~{\sc xvii }$2p$+sat, S~{\sc xv }$4p$      \\
    4 & 4.120~-~4.230 & cont. + S~{\sc xv }sat.                              \\
    5 & 4.340~-~4.430 & cont. + S~{\sc xv }$3p$+sat                          \\
    6 & 4.430~-~4.520 & cont. + Cl~{\sc xvi }$2p$+sat                        \\
    7 & 4.680~-~4.750 & cont. + S~{\sc xvi }$2p$, + Si~{\sc xiv }$8p$        \\
    8 & 4.750~-~4.800 & cont. + Si~{\sc xiv }$6p$                            \\
    9 & 5.000~-~5.150 & S~{\sc xv }$2p$+sat + cont.                          \\
   10 & 5.220~-~5.320 & Si~{\sc xiii }$5p$+sat.  + cont.                     \\
   11 & 5.320~-~5.470 & Si~{\sc xiii }$4p$+sat.  + cont.                     \\
   12 & 5.475~-~5.640 & Si~{\sc xii } sat. + cont.                           \\
   13 & 5.640~-~5.715 & Si~{\sc xiii }$3p$ + cont.                           \\
   14 & 5.715~-~5.850 & Si~{\sc xii } sat. + cont.                           \\
   15 & 5.900~-~5.950 &  continuum                                           \\
\hline                                   %inserts single line
\end{tabular}
\vspace{-5mm}
\end{table}

%%%%%%%%%%%%%%%%%%%%%%%%%%%%%%%%%%%%%%%

The analysis proceeds in two steps. In the first (called AbuOpt), optimized values for the abundances of elements that make large contributions to the line features in RESIK spectra are found. We selected the input data which consisted of observed fluxes in 15 narrow spectral intervals in RESIK spectra for each of the 19 time intervals ($a$ to $s$) shown in Figure~\ref{norm_RHESSI_GOES_lcs}; most include strong emission lines while a few are almost entirely continuum radiation (the sum of free--free and free--bound). The wavelength ranges of the intervals and details of principal emission features included (lines or continuum) are specified in Table~\ref{table:1}. The main lines are those due to H-like or He-like Si, S, and Ar and associated dielectronic satellite lines, with the He-like K lines generally appearing as weak (the He-like Cl lines are much weaker still and have practically no effect on the analysis). In the table, we use the notation ``$2p$" for the transition $1s^2\,^1S_0 - 1s2p\,^1P_1$ in He-like ions and ``$2p$" for Ly-$\alpha$ in H-like ions; ``sat." for dielectronic satellite lines; and ``cont." for continuum. We then applied the Withbroe--Sylwester DEM inversion method to these input data to achieve an optimum fit between the observed and calculated fluxes. This requires the evaluation of the emission functions $f_i(T)$ for these intervals -- lines plus continuum or continuum alone -- which can be calculated from the {\sc chianti} (v.~7.0) code. These depend on temperature and element abundance (Equation~\ref{lineflux}). We investigated the effect of varying the abundances of Si, S, Ar, and K. To do this, and to avoid a time-consuming calculation, we pre-calculated a grid of theoretical spectra with temperatures in the range 1--100~MK (101 steps in equal intervals of log~$T$) and 21 values of the abundances of four elements (Si, S, Ar, and K) ranging from zero to 16 times the {\sc chianti} ``coronal" value. With one exception, the other element abundances, which affect the free--bound continua only, were kept at their coronal values as specified in {\sc chianti}; the effect of varying these abundances is relatively minor -- see the analysis of \cite{phi10b}. The exception is the element Cl which has very low abundance; for this we took our abundance estimate $A({\rm Cl}) = 5.75 \pm 0.26$ (from \cite{bsyl11} based on an isothermal analysis of \ion{Cl}{16} lines in RESIK channel~3). There are, as a result of this calculation, 84 ($4 \times 21$) pre-calculated spectra for each of 101 temperatures in the RESIK spectral range (the spectral resolution was chosen to be 0.001~\AA).

\begin{figure}
\epsscale{.60}
\plotone{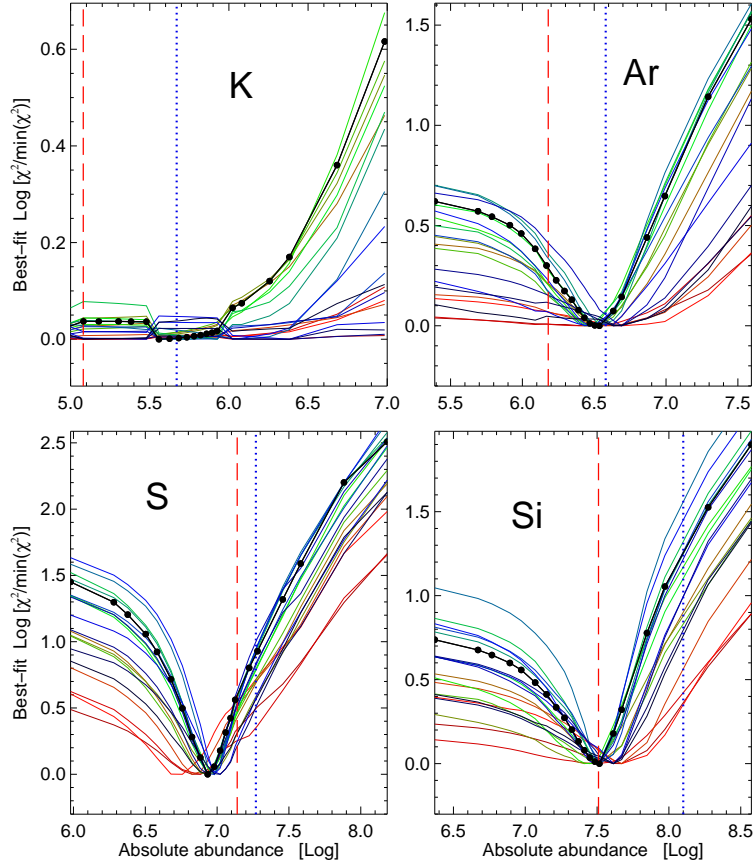}
\caption{Plot of the quality of the fit, expressed as the ratio of normalized $\chi^2$ to the minimum value of $\chi^2$, of the observed and theory fluxes for all 19 time intervals during the flare (colored to distinguish them: blue for rise phase spectra, red, orange, or green for decay phase) as a function of assumed element abundance for K, Ar, S and Si. The black curve with dots is derived from the average spectrum over the 19 time intervals.  The vertical dotted blue lines correspond to the {\sc chianti} ``coronal" abundance \citep{fel92b}, and the dashed red lines to the {\sc chianti} ``photospheric" abundances \citep{asp09}.}
\label{abund_chi2}
\end{figure}

With these calculated spectra, we ran the Withbroe-Sylwester DEM method on the input data consisting of fluxes in the 15 narrow spectral intervals, the DEM ($\varphi(T)$) function varying freely for each of the 21 abundance values for Si, S, Ar, and K. After 1000 iterations, the value of normalized $\chi^2$ was obtained describing the difference between the measured and fitted fluxes in terms of measurement uncertainties (assuming Poissonian statistics of the photon counts). Figure~\ref{abund_chi2} shows examples for Si, S, Ar, and K for the 19 time intervals (represented by curves with different  colors) and the spectrum averaged over all 19 intervals (black curve with dots). Clear minima in the value of $\chi^2 / {\rm min} (\chi^2)$ as a function of element abundance are apparent for Ar, S, and Si; the K abundance is too small to have any effect on the value of $\chi^2$ below a threshold corresponding to a potassium abundance of $10^{-6}$ of the hydrogen abundance. We interpret the plots to mean that the abundance corresponding to the minimum in $\chi^2$ is the optimum one, i.e. for which the agreement between the observed set of spectral fluxes and the theory is the best.

\begin{figure}
\epsscale{0.70}
\plotone{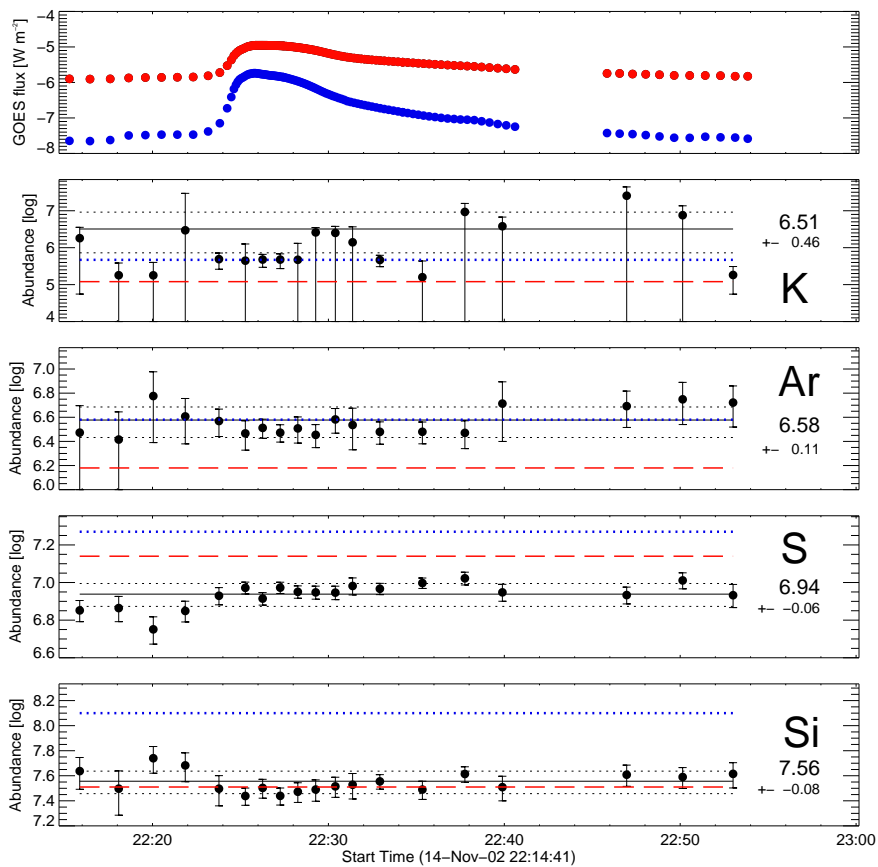}
\caption{The time variations of derived absolute abundance of K, Ar, S and Si. The abundances were determined using the abundance--optimization (AbuOpt) approach described in the text. The error bars on abundance determination are based on the results presented in Figure~\ref{abund_chi2} and correspond to the range of values for ${\rm min}(\chi^2)+1.0$. Thin black horizontal lines represent time averaged values of elemental abundances together with their RMS error bands (dotted horizontal lines). In the top panel, the $\emph{GOES}$  light curves integrated over the times of individual RESIK spectra collection are shown. The horizontal blue dotted lines correspond to ``coronal" abundances (\citet{fel92b}), while the dashed red lines correspond to  ``photospheric" abundances (\citet{asp09}).}
\label{t_varns_abunds}
\end{figure}

In Figure~\ref{t_varns_abunds} we plot derived optimum abundance values against time for Si, S, Ar, and K. Uncertainties in the abundance determinations are assigned that correspond to the abundance range defined by ${\rm min}(\chi^2) + 1$ \citep{bev03}. There is little evidence for time-changing abundances apart from a slight tendency of $A({\rm S})$ to be a little smaller at earlier times. The abundances (linearly averaged over the 19 time intervals) obtained are: $A({\rm K}) = 6.51 \pm 0.46$, $A({\rm Ar}) = 6.58 \pm 0.11$, $A({\rm S}) = 6.94 \pm 0.06$, and $A({\rm Si}) = 7.56 \pm 0.08$ (abundances are expressed on a logarithmic scale with $A({\rm H}) = 12$). As a result, we assume from hereon that the Si, S, Ar, and K abundances are time- and temperature-independent, at least for the flare considered here. In future analysis, this assumption may be relaxed to account for possible abundance differences with temperature.

%%%%%%%%%%%%%%%%%%%%%%%%%%%%%%%  Figure 4   %%%%%%%%%%%%%%%%%%%%%%%%%%%%%%%%%%%%%%%%%%%%%%%%%%%%%%%
%%%%%%%%%%%%%%%%%%%%%%%%%%%%%%%%%%%  Figure 4  %%%%%%%%%%%%%%%%%%%%%%%%%%%%%%%%%%%%%%%%%%%%%%%%

As indicated earlier, most of our previous analyses of RESIK flare spectra were based on an isothermal assumption, with the emitting temperature and emission measure taken from {\em GOES} ($T_{\rm GOES}$, $EM_{\rm GOES}$). For the Ar abundance, determined from the prominent \ion{Ar}{17} lines in channel~2, a very tight distribution of observed points given by the ``$G(T)$" plot, i.e. line flux divided by $EM_{\rm GOES}$ plotted against $T_{\rm GOES}$, was obtained, resulting in an abundance determination with small uncertainty: $A({\rm Ar}) = 6.45 \pm 0.06$ (Figure~2 of \citet{jsyl10b}). This value is in close agreement with other determinations from solar proxies (e.g. \ion{H}{2} regions, Jupiter's atmosphere) and is considered by us to be fairly definitive. The value obtained in this work is slightly higher. It also has larger uncertainty which would indicate {\em prima facie} that for this particular flare the isothermal abundance interpretation is more likely to be correct. But as indicated in Section~1, it appears, from the agreement of the observed points with the calculated $G(T)$ plot shown by \citet{jsyl10b}, that the characteristic temperature of the plasma emitting the \ion{Ar}{17} lines is well described by the temperature estimated from the two {\em GOES} channels.  We therefore interpret this to mean that the isothermal abundance determination is to be preferred. For potassium, the estimated abundance from this work is a factor 4.5 higher than that from our previous determination ($A({\rm K}) = 5.86 \pm 0.23$: \citet{jsyl10a}) assuming isothermal emission and has larger uncertainty. The larger uncertainties again suggest that the value from our earlier isothermal analysis to be preferred; the similar temperature of formation of the \ion{K}{18} lines to that of the \ion{Ar}{17} lines suggests that the {\em GOES} temperature accurately describes the \ion{K}{18} line emission also.

For the lower-temperature S ions (both H-like and He-like ions were considered), the isothermal analysis of RESIK spectra by \citet{jsyl12} was not particularly satisfactory, with large scatter of points on the $G(T)$ plot, even though there are several strong lines with which the analysis is possible. \citet{jsyl12} determined $A({\rm S}) = 7.16 \pm 0.17$, i.e. higher than the presently determined value ($A({\rm S}) = 6.94 \pm 0.06$) and with larger uncertainty. A re-analysis of the SOL2002-11-14T22:26 flare spectra alone on an isothermal assumption gives $A({\rm S}) = 7.15 \pm 0.10$ (\ion{S}{15} $w$ and nearby lines at 5.04~\AA) and $A({\rm S}) = 7.08 \pm 0.18$ (\ion{S}{15} $w4$ line at 4.08~\AA). Our interpretation here is that the less satisfactory agreement of the observed points with the calculated $G(T)$ function is due to the fact that the \ion{S}{15} emission functions have characteristic temperatures that are less than those derived from the ratio of the {\em GOES} channels. In this case, a DEM analysis of the emitting flare plasma gives a more reliable abundance determination. In summary, our preferred value for the S abundance for this flare is the one determined in the present analysis, viz. $A({\rm S}) = 6.94 \pm 0.06$.

The argument for the S abundance determination holds {\em a fortiori} for the Si abundance, since the lines on which the determinations are principally made, He-like and H-like Si (\ion{Si}{13} and \ion{Si}{14}), have lower characteristic temperatures than the S lines. From an isothermal assumption \citep{bsyl13}, we determined $A({\rm Si}) = 7.89 \pm 0.13$ (\ion{Si}{13} lines), and $A({\rm Si}) = 7.93 \pm 0.21$ (\ion{Si}{14} lines), both estimates being about a factor 2 more than is estimated from the present work ($A({\rm Si}) = 7.56 \pm 0.08$) and having larger uncertainties. In this case, therefore, our preferred value for the Si abundance is the one given by this analysis. (Note that for comparison with recent abundance estimates for RS CVn stars \citep{hue13a}, values of S and Si abundances were used from a DEM analysis of a flare on 2002 December~26 with the isothermal values of Ar and K for the same flare.)

An illustration of the foregoing argument with reference to the $G(T)$ functions for the principal RESIK lines is provided by a plot that we constructed  of RESIK and {\em RHESSI} spectra during a flare against photon energy together with a theoretical spectrum from {\sc chianti} calculated with {\em GOES} values of temperature and emission measure estimated at the time of the spectrum. A very close agreement of the {\sc chianti} spectrum with the RESIK channels 1 and 2 spectra was found. However, the temperature of the {\sc chianti} spectrum was too low to describe the {\em RHESSI} spectrum but on the other hand too high to describe the RESIK spectrum in channels 3 and 4. As RESIK channels 1 and 2 include the \ion{Ar}{17} and \ion{K}{18} lines, it appears that these lines are well described by {\em GOES} temperature and emission measure but the S and Si lines in RESIK channels 3 and 4 are not; for these lines a lower-temperature component is needed to fit their fluxes.

%%%%%%%%%%%%%%%%%%%%%%%%%%%%%%%  Figure 5   %%%%%%%%%%%%%%%%%%%%%%%%%%%%%%%%%%%%%%%%%%%%%%%%%%%%%%%
%%%%%%%%%%%%%%%%%%%%%%%%%%%%%%%%%%%  Figure 5  %%%%%%%%%%%%%%%%%%%%%%%%%%%%%%%%%%%%%%%%%%%%%%%%

\begin{figure}
\epsscale{.70}
\plotone{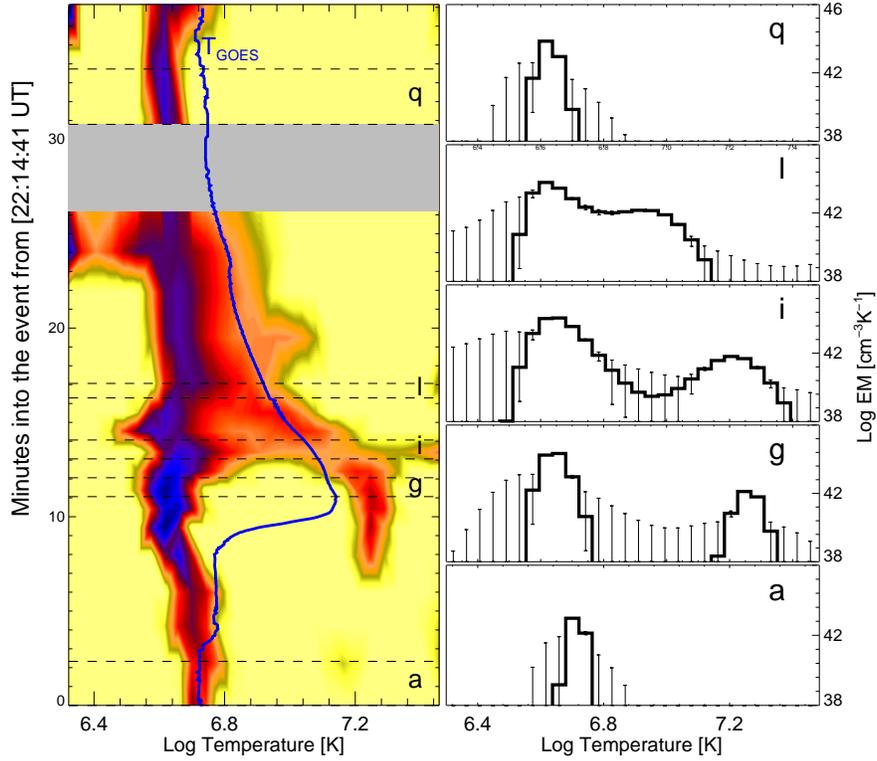}
\caption{(Left:) Contour plot of the differential emission measure during the SOL2002-11-14T22:26 flare, darker colors indicating greater emission measure. The horizontal scale is the logarithm of temperature, and time increases upwards, measured from 22:14:41~UT. Horizontal dotted lines define the time intervals $a$, $g$, $i$, $l$, and $q$ (see Figure~\ref{norm_RHESSI_GOES_lcs}) and the smooth curve running from top to bottom is the temperature derived from the ratio of the two {\em GOES} channels on an isothermal assumption. (Right:) Emission measure distributions for the intervals indicated in the left plot, derived from the Withbroe--Sylwester routine. Vertical error bars indicate uncertainties. A cooler (temperature $\sim 4-5$~MK) component is present over all the time interval shown, with hotter component ($\sim 18$~MK) at the peak of the {\em GOES} light curve. }
\label{2D_DEM}
\end{figure}

With optimized averaged abundances for S and Si and the isothermal abundance determinations for Ar and K, the DEM distribution for the 19 time intervals of the flare could now be calculated. Abundances of elements other than these were taken from the {\sc chianti} ``coronal" abundances of \citet{fel92a}, except for Cl as already described. With the ionization fractions of \citet{bry09}, the Withbroe--Sylwester procedure was then used for the DEM inversion, the convergence continuing until iteration $10\,000$. The uncertainties of the inversion  were determined from 100 Monte Carlo runs, where the input line fluxes for every time step were randomly perturbed with corresponding statistical uncertainties. The evolutionary changes of  the DEM are indicated in the left and right panels of Figure~\ref{2D_DEM}. The left panel shows the DEM distribution as a contour plot, while calculated DEMs for selected times during the flare (intervals $a$, $g$, $i$, $l$, $q$ in Figure~\ref{norm_RHESSI_GOES_lcs}) are shown in the right panel. It is evident that for all times the bulk of the emitting plasma has a temperature of $3-6$~MK contributing to a $\emph{cooler}$ component. A {\em hotter} component is present near the flare's maximum phase with temperature of $\sim 16-21$~MK as is clear from Figure~\ref{2D_DEM}, but the emission measure is always more than two orders of magnitude smaller. For comparison, the temperature derived from the flux ratio of the two bands of {\em GOES}, assuming isothermal emission, is shown on this plot as the curve running from top to bottom; the {\em GOES} temperature is generally a little higher than the cooler component indicated by the DEM analysis except at the flare peak when it does not quite attain the temperature of the hotter component. The temperature of the cooler component is perhaps somewhat larger than that typical of non-flaring regions, but the {\em GOES} emission in the previous {\em CORONAS-F} spacecraft orbit (at about the C1 level) indicates a temperature of about 5~MK, a reflection of the presence of several non-flaring active regions on the Sun at the time. A DEM analysis of the RESIK emission during the previous orbit in fact shows the bulk of the emission to have a temperature of $\sim 2.8$~MK.

%%%%%%%%%%%%%%%%%%%%%%%%%%%%%%%  Figure 6   %%%%%%%%%%%%%%%%%%%%%%%%%%%%%%%%%%%%%%%%%%%%%%%%%%%%%%%
%%%%%%%%%%%%%%%%%%%%%%%%%%%%%%%%%%%  Figure 6  %%%%%%%%%%%%%%%%%%%%%%%%%%%%%%%%%%%%%%%%%%%%%%%%

\begin{figure}
\epsscale{.60}
\plotone{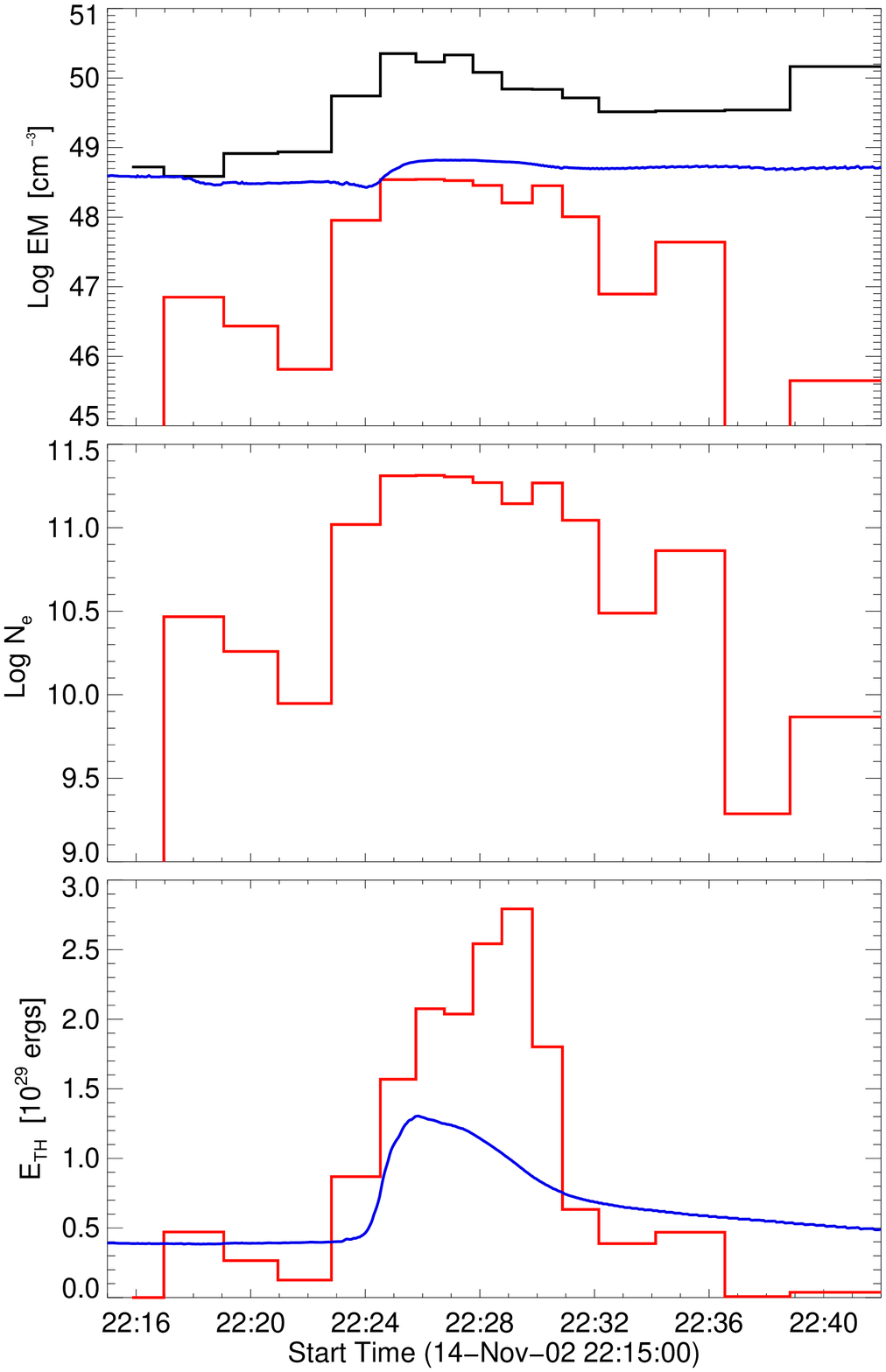}
\caption{The time evolution of (top) the total emission measure for the cooler ($T<9$~MK, in black) and hotter ($T>9$~MK in red) plasma (the blue solid line is the emission measure $EM_{\rm GOES}$ from the flux ratio of the {\em GOES} bands); (center) electron densities derived from the emission measure of the hotter component and average size of the {\em RHESSI} images; (bottom) thermal energy $E_{\rm th}$, defined by Equation~(\ref{thermal_measure}) (the blue curve is $E_{\rm th}$ as deduced from {\em GOES}).}
\label{EM_Ne_ThM}
\end{figure}

The total emission measures of the cooler and hotter components indicated in Figure~\ref{2D_DEM} (left panel) were evaluated and plotted in Figure~\ref{EM_Ne_ThM} (top panel). The emission measures of both components evolve with time, reaching a maximum in each case at about the time of the {\em GOES} maximum emission (22:26~UT), with that of the cooler component larger by a factor of about 100. The decrease of the emission measure of both components is somewhat faster than that of $EM_{\rm GOES}$ evaluated from the flux ratio of the two {\em GOES} bands.

%%%%%%%%%%%%%%%%%%%%%%%%%%%%%%%  Figure 7 %%%%%%%%%%%%%%%%%%%%%%%%%%%%%%%%%%%%%%%%%%%%%
%%%%%%%%%%%%%%%%%%%%%%%%%%%%%%%  Figure 7  %%%%%%%%%%%%%%%%%%%%%%%%%%%%%%%%%%%%%%%%%%%%%

The temperature of the cooler component is low enough that it is unlikely to contribute significantly to the 6~--~7~keV emission seen in the {\em RHESSI} images shown in Figure~\ref{RHESSI_images}. If we interpret the {\em RHESSI} emission to be due mainly to the hotter component seen in RESIK spectra, we may obtain densities and other information about the flaring plasma. A series of approximately 60 {\em RHESSI} 6~--~7~keV images of the (SOL2002-11-14T22:26) flare shows that the main emission component is a confined feature likely to be located at the top of a small (few thousand km) loop. The size of this component is constant to within 25\% or so, and as indicated earlier at the 50\% iso-contour level the extent is about 5400 km. An assumed spherical shape for this leads to a volume of $8.2 \times 10^{25}$~cm$^3$. Combined with the peak emission measure of the hotter component ($3 \times 10^{48}$~cm$^{-3}$), this gives an electron density $N_e = 2.6 \times 10^{11}$~cm$^{-3}$. The detailed time variations are indicated in Figure~\ref{EM_Ne_ThM} (center panel). Estimates of the thermal energy $E_{\rm th}$,

% Eq. 3
\begin{equation}
E_{\rm th}\mid_{N_e={\rm const}} = 3k_B \frac{\int{ T\varphi(T) dT}} {\sqrt {\int{\varphi(T)dT}}}\sqrt{V}
\label{thermal_measure}
\end{equation}

\noindent where $k_B$ is Boltzmann's constant, presented in Figure~\ref{EM_Ne_ThM} (bottom panel), show that $E_{\rm th}$ reaches a maximum of $\sim 3 \times 10^{29}$~erg, rather typical for a medium-class flare such as the one under discussion.

% Section 4
\section{Discussion and Conclusions}

One of the primary intentions of this analysis of RESIK spectra and other data for the M1 flare under discussion has been to test a calculation procedure for obtaining differential emission measure using spectral fluxes from the RESIK instrument on {\em CORONAS-F}. Data from this well-calibrated instrument have been used in the past to derive abundances of elements whose lines occur in the RESIK X-ray range (3.4--6.1~\AA), viz. Si, S, Ar, K, and Cl. For flare data we have previously used the approximation that the emitting plasma is isothermal with temperature given by the flux ratio of the two bands of {\em GOES}. This appears to be a good assumption for the case of Ar, K, and Cl abundance determinations, which particularly in the case of Ar have small uncertainties and agree well with abundance determinations from other, unrelated, methods. Previous determinations of the Si and S abundances from RESIK spectra using the same procedure appear to be less accurate: the estimates have larger uncertainties, and the characteristic temperatures of the emission functions of H-like and He-like ions of these elements are significantly less than that from the flux ratio of the {\em GOES} bands. Here we use a procedure (called AbuOpt) in which the abundances of Si, S, Ar, and K are first optimized using the maximum-likelihood, Bayesian Withbroe--Sylwester inversion technique for obtaining differential emission measure from line fluxes is used. The Withbroe--Sylwester routine was then run with optimized abundances to obtain the time evolution of the DEM. The result (shown as a time sequence in Figure~\ref{2D_DEM}) is a DEM distribution with well-defined cooler ($\sim 3-6$~MK) and hotter ($16-21$~MK) components. If the hotter component is assumed to describe the emission seen by {\em RHESSI}, the spatial dimensions combined with the total emission measure of the hotter component lead to estimates of electron density and thermal energy. Compared with recent estimates of $N_e$ from extreme ultraviolet line ratios, these represent lower limits, but give an indication of the physical characteristics of the flaring plasma.

Values of element abundances derived from both an isothermal assumption and from the AbuOpt method described here were discussed in \S 3. It appears that our previous estimates for Ar and K based on an isothermal assumption are reliable, having smaller uncertainties, and are preferred values, but for S and Si, the abundance estimates from the present work -- $A({\rm S}) = 6.94 \pm 0.06$ and $A({\rm Si}) = 7.56 \pm 0.08$ -- are to be preferred for the flare analyzed here.

Our Si abundance estimate is significantly lower than our estimates from an isothermal analysis of RESIK flare spectra \citep{bsyl13}, viz. $A({\rm Si}) = 7.93 \pm 0.21$ from the \ion{Si}{14} Ly-$\beta$ line at 5.217~\AA\ and $A({\rm Si}) = 7.89 \pm 0.13$ from the \ion{Si}{13} $w3$ line at 5.688~\AA. There are very few determinations of the Si abundance from X-ray flare spectra that our values can be compared with, the only reliable one being that of \cite{vec81} who used the {\em OSO-8} graphite crystal spectrometer to derive $A({\rm Si}) = 7.73 ^{+ 0.19}_{- 0.35}$ from the \ion{Si}{13} $w3$ line and $7.62 ^{+ 0.13}_{- 0.34}$ from the \ion{Si}{14} Ly-$\alpha$ line. An isothermal assumption was used, with the temperature derived from the slope of the nearby continuum which is evident in the spectra from this instrument. Clearly, our present estimates are nearer to the \cite{vec81} values than those from our isothermal analysis. Photospheric abundance estimates \citep{asp09} from 3-D and 1-D LTE analyses, one with non-LTE corrections, give $A({\rm Si}) = 7.51 \pm 0.04$, so abundance estimates from our present X-ray flare result and from the \ion{Si}{14} result of \cite{vec81} are not significantly larger (factor $1.12 \pm 0.2$) than photospheric despite the enhancement expected from the low FIP value for Si (8.15~eV); at any rate the enhancement is less than the factor 4 indicated by \cite{fel92a} and \cite{fel92b}.

The S abundance estimate obtained here is only 0.03 different from the determination of \cite{vec81} from the {\em OSO-8} instrument: they give $A({\rm S}) = 6.91^{+ 0.13}_{- 0.19}$ determined from the intense \ion{S}{15} $w, y, z$ lines at $\sim 5$~\AA. Again, our result and the \cite{vec81} result are unexpected on the standard FIP picture for this element which has a FIP (10.4~eV) marginally considered to be high (i.e. more than 10~eV). Recent photospheric estimates range from $A(S) = 7.12 \pm 0.03$ \citep{asp09} to $7.15 \pm (0.01)_{\rm stat} \pm (0.05)_{\rm syst}$ \citep{caf11}. If the uncertainty estimates for all these determinations are to be considered literally, there would appear to be a slight inverse FIP effect for S, i.e. our flare abundance estimate is $0.6 \pm 0.1$ times photospheric. An inverse FIP effect is difficult to reconcile with theoretical models (e.g. \cite{hen98}) with the exception of that based on ponderomotive forces associated with propagating Alfv\'{e}n waves proposed by \cite{lam04,lam09,lam12}. An inverse FIP effect, which is observed in cool main sequence stars \citep{wood12}, can possibly be explained by waves propagating upwards from the chromosphere and reflecting back down as opposed to propagating downwards from the corona and reflecting back upwards.

Extreme-ultraviolet spectral lines observed by the Extreme-ultraviolet Imaging Spectrometer (EIS) on {\em Hinode}  have recently been used to give Si/S abundance ratios in active regions and other non-flaring features from the intensity ratio of \ion{Si}{10} 258.37~\AA\ line to the \ion{S}{10} 264.23~\AA\ line. The ratio was determined by \cite{bro11} to range from 2.5 to 4.1 (average 3.4) and by \cite{bak13} from 2.5 to 3 for an ``anemone" active region within a coronal hole but up to more than 4 for established active regions. The Si/S abundance ratio from the estimates given here is $4.16^{5.25}_{3.31}$, so on the basis of the EIS results is typical of an established active region. The 2002 November~14 flare occurred in an active region that had recently appeared on the Sun's south-east limb, so its history is not well determined; all one can say is that it was  flare-prolific on November~14, with a complex magnetic geometry.

The electron densities in this analysis are typical of those estimated from a combination of image data and volume emission measures, but are less than those from density-sensitive spectral line ratios. These are in short supply in the X-ray region, but a number of \ion{Fe}{21} lines in the extreme ultraviolet spectrum seen with the Extreme Ultraviolet Variability Experiment (EVE) instrument on {\em Solar Dynamics Explorer} have recently been identified, enabling electron density to be found as a function of time in two {\em GOES} class X flares \citep{mil12}. These indicate densities of up to $10^{12}$~cm$^{-3}$, a factor 5 higher than the M1.0 flare discussed here. Our estimates of both $N_e$ and the thermal energy are thus likely to be lower limits with undetermined filling factors, perhaps of order 1/25.

We plan to use the AbuOpt approach together with the Withbroe-Sylwester DEM analysis method to study element abundances and the thermodynamics of several other flares observed by RESIK to see whether the reduced Si and S abundances found here still hold, and whether any variations in element abundance are related to flare characteristics such as {\em GOES} class.

\acknowledgments

We acknowledge financial support from the Polish National Science Centre grant number 2011/01/B/ST9/05861 and the European Commissions grant FP7/2007-2013: eHEROES, Project No. 284461. The data analysis was performed using the {\sc chianti} atomic code, a collaborative project involving George Mason University, University of Michigan (USA) and University of Cambridge (UK). The launch and operation of RESIK was possible thanks to generosity of the Russian team of scientists and engineers from the IZMIRAN Institute led by Professor V.~D. Kuznetsov.

%% Using .bib file
\bibliography{RESIK}{}
\bibliographystyle{apj}

\end{document}